\documentclass[apj]{emulateapj}
\usepackage{graphics}
\usepackage{amssymb,amsmath}

\shorttitle{CHANDRA STUDY OF NGC~326}
\shortauthors{HODGES-KLUCK \& REYNOLDS}

\slugcomment{Accepted by ApJ}

\begin{document}

\title{A Chandra Study of the Radio Galaxy NGC~326: \\Wings, Outburst History, and AGN Feedback}

\author{Edmund~J.~Hodges-Kluck$^{1}$ \& Christopher~S.~Reynolds$^{2,3}$}
\altaffiltext{1}{Department of Astronomy, University of Michigan, Ann Arbor,
MI, 48109}
\altaffiltext{2}{Department of Astronomy, University of Maryland, College
Park, MD 20742}
\altaffiltext{3}{Joint Space Science Institute (JSI), University of Maryland, College
Park, MD 20742}
\email{hodgeskl@umich.edu}

\begin{abstract}
NGC~326 is one of the most prominent X- or Z-shaped radio galaxies (XRGs/ZRGs) and
has been the subject of several studies attempting to explain its
morphology through either fluid motions or reorientation of the jet
axis.  We examine a 100~ks \textit{Chandra X-ray Observatory} exposure
and find several features associated with the radio galaxy: a
high-temperature front that may indicate a shock, high-temperature
knots around the rim of the radio emission, and a cavity associated
with the eastern wing of the radio galaxy.  A reasonable
interpretation of these features in light of the radio data allows us
to reconstruct the history of the AGN outbursts.  The active outburst
was likely once a powerful radio source which has since decayed, and
circumstantial evidence favors reorientation as the means to produce
the wings.  Because of the obvious interaction between the radio
galaxy and the ICM and the wide separation between the active lobes
and wings, we conclude that XRGs are excellent sources
in which to study AGN feedback in galaxy groups by measuring the
heating rates associated with both active and passive heating
mechanisms.  
\end{abstract}


\section{Introduction}

A small fraction of double-lobed extragalactic radio sources have
long, centro-symmetric tails of radio emission extending from
the nucleus or primary lobes in a very different direction, thereby
producing ``X'' or ``Z''-shaped radio galaxies (XRGs/ZRGs).  The tails, or
``wings'', are faint, do not currently harbor a relativistic jet, and
can be substantially longer (in projection) than the primary lobes.
They have elicited considerable interest due to the suggestion by
\citet{merritt02} that they result from a recent supermassive black
hole (SMBH) merger that reoriented the jet, but their origin is uncertain.  It has also been
suggested that the wings result from the deflection of lobe plasma by
pressure fronts in the intergalactic or intracluster medium (IGM/ICM)
\citep{leahy84,worrall95,capetti02,gopal03,zier05,hodges-kluck10a}.  Neither of these
hypotheses have yet been proven, even in a single source. 

\begin{figure*}[t]
\begin{center}
\includegraphics[width=0.42\textwidth,angle=-90]{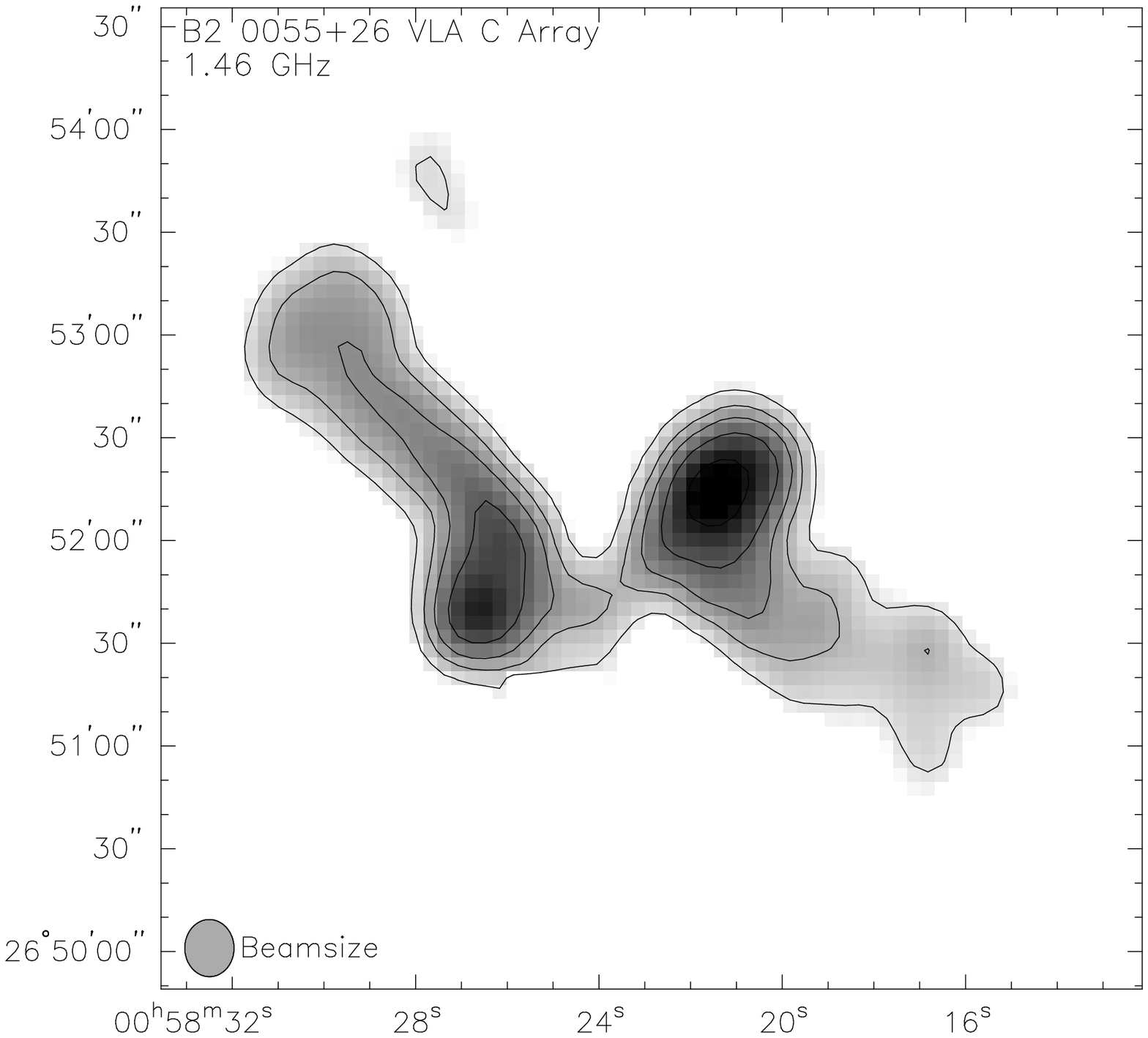}
\includegraphics[width=0.42\textwidth,angle=-90]{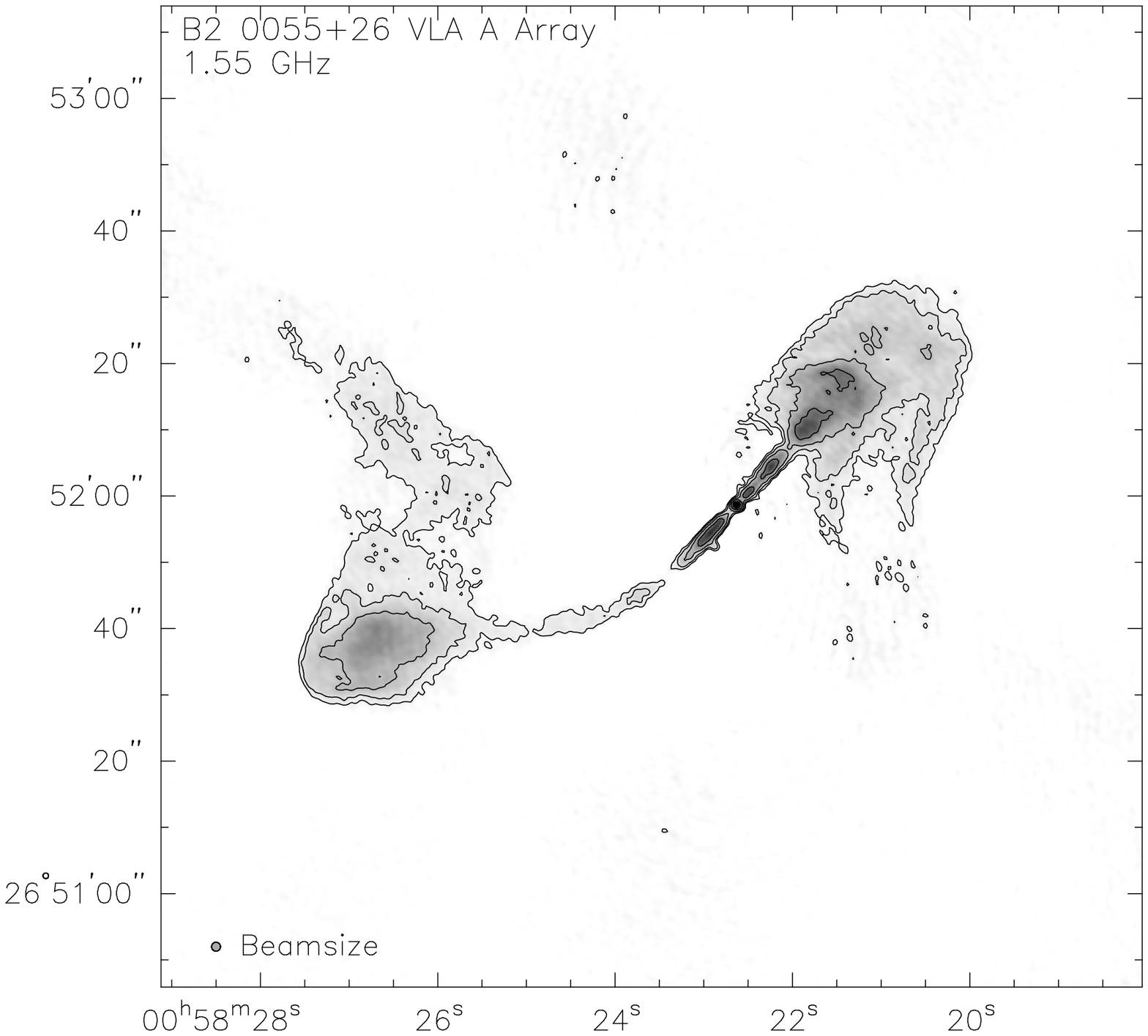}
\caption{\textit{Left:} 1.46 GHz VLA C Array image.  The contour levels are
1, 2, 4, 8, 16, 32, 64, and 128~mJy/beam.  The restoring beamsize is 
$16.^{\prime\prime}7\times 14.^{\prime\prime}4$.
\textit{Right:} 1.55 GHz VLA A Array.  The contour levels are 0.1, 0.2,
0.4, 0.8, 1.6, and 3.2~mJy/beam.  The restoring beamsize is 
$1.^{\prime\prime}35\times 1.^{\prime\prime}15$.  Note that the images are
on different size scales, and the full extent of the wings is not seen at
right.
}
\label{radio_maps}
\end{center}
\end{figure*}

B2~0055+26 is a dramatic XRG/ZRG, with wings longer (in projection) than
the primary lobes (Figure~\ref{radio_maps}).  The wings are highly
collimated over $\sim$100~kpc.  Its host galaxy is the northern
component of the dumbbell NGC~326 \citep{colla75} and the entire system
resides in the galaxy cluster Zw~0056.9+2636 \citep{zwicky68}.
Hereafter, we adopt the convention in the literature and refer to the
whole system as NGC~326.  The
cluster atmosphere has been characterized by \citet[][hereafter,
  W95]{worrall95} on large scales using ROSAT, and has an
average temperature of $kT \sim 2$~keV.  The cluster atmosphere is
substantially asymmetric (Figure~\ref{rosat}), suggesting that it is
not in hydrostatic equilibrium and may be made up of two or more
merging subclusters or that it is a composite of two or three poorer
clusters seen in projection (W95).  A detailed multifrequency study of
the radio emission measured the spectral index $\alpha$ as a function
of position in the lobes and wings, finding spectral steepening in the
wings and away from the ends of the jets \citep[][hereafter M01]{murgia01}.

Owing to its dramatic structure (Figure~\ref{radio_maps}), NGC~326 is
frequently used as the prototypical XRG/ZRG in formation models
purporting to explain the X-shaped morphology.  \citet{ekers78}, who
discovered the wings but not their true length, attributed them to
ongoing regular precession of the relativistic jet inflating the
primary lobes.  Later, \citet{merritt02} argued that the striking
X-shaped structure necessitated rapid secular precession.  Noting that
black holes are difficult to reorient via external forces other than
accretion disk instabilities, \citet{merritt02} suggested instead that
the X-shaped structure is the result of the final coalescence of a
SMBH binary.  Supposing that the components of the binary have
misaligned angular momenta, the final spin axis (and jet axis) will be
nearly instantaneously reoriented.  \citet{wirth82} also argued in
favor of a single rapid jet reorientation caused by a close encounter
between the constituents of the dumbbell galaxy.

Based on the large-scale, hot, asymmetric cluster atmosphere, W95
found that the buoyancy of the lobe material could explain the wings
provided that they were driven by rapid backflows expected near the
terminal shocks of the radio jet heads.  This scenario is
fundamentally different from those above in that it presumes a stable
jet axis.  In support of the backflow hypothesis, we
\citep{hodges-kluck10a} reported a local asymmetric enhancement of the
hot gas in a preliminary analysis of an archival 100~ks \textit{Chandra
  X-ray Observatory} (CXO) image on the
scale of $\sim$50--75~kpc where the radio beam is aligned with the
major axis and the wings with the minor axis.  This is also consistent
with the expectation of \citet{capetti02} that X-shaped sources are
aware of the geometry of their environments.  However, it is not clear
how the wings would grow so long in these models. 

In this paper, we carefully consider the high-resolution
CXO data in order to shed light on the history of
NGC~326, evalauting possible hydrodynamic hypotheses and
characterizing the interaction between the radio galaxy and its
environment, which could not be resolved in the W95 study.  
 NGC~326 is located at a redshift $z = 0.0474$ where 
$1^{\prime\prime} = 0.92$~kpc in the \textit{Wilkinson Microwave Anisotropy
Probe} cosmology \citep[$H_0 = 71$~km~s$^{-1}$~Mpc$^{-1}$, $\Omega_{\Lambda} = 0.73$, and 
$\Omega_m = 0.27$;][]{spergel07}.  We use a Galactic photoelectric absorption
in all spectral fitting of $N_H = 5.69\times 10^{20}$~cm$^{-2}$
\citep{kalberla05}.  The total extent of the cluster atmosphere in the ROSAT 
image is about 20$^{\prime}$ on a side (1.1~Mpc).

The remainder of the paper is organized as follows: in Section~2 we
describe the CXO observation, data, and basic features, then in
Section~3 we use the X-ray and radio data to derive quantities of
interest related to the radio galaxy.  In Section~4 we use these data
to reconstruct the history of the radio galaxy and argue that NGC~326,
as well as other XRGs, are interesting sources with regard to feedback
from active galactic nuclei (AGNs).  We summarize our results in
Section~5.  

\section{Observations}

NGC 326 was observed on 2006-09-02 for 94.4~ks by CXO 
using the Advanced CCD Imaging Spectrometer\footnote{See http://cxc.harvard.edu/proposer/POG/pdf/ACIS.pdf} 
(ACIS) with the dumbbell galaxies located at the nominal aimpoint on the
S3 chip (obs. ID 6830).  We previously published these dating looking only at gross
morphology \citep{hodges-kluck10a}.  
The data were reduced with the \textit{Chandra
Interactive Analysis of Observations} (CIAO v4.0) software, starting from the
level=1 file rather than using the pipeline level=2 file.  No background
flares in the $0.3-10$~keV bandpass were detected in the lightcurve of 
low surface brightness regions on the S3 chip or empty regions on other chips
(the entire S3 chip is covered by some cluster 
emission).  For spectral fitting we used XSPEC v12.7.0 \citep{arnaud96}, with the 
{\tt apec} model as our thermal model \citep{smith01}.  Images 
produced from the reduced events file were exposure corrected using an exposure
map generated in CIAO, and questionable pixels at the very edge of the chip 
which result from this process are excised from our analysis.  For analysis
of the hot cluster atmosphere, point sources are excised from the image using
the CIAO {\tt wavdetect} tool.  We also exclude the resolved interstellar
media (ISM) of the dumbbell galaxies.  Since the cluster emission extends north
onto the S2 chip, we follow an analogous procedure for these data. 

In addition to the CXO data, we used archival NRAO\footnote{The National Radio Astronomy Observatory 
is a facility of the National Science Foundation operated under cooperative
agreement by Associated Universities, Inc.} \textit{Very Large
Array} (VLA) radio maps presented in M01.  These include
maps at 1.4, 1.6, 4.8, 8.5, and 14.9~GHz produced with the NRAO AIPS 
software.  Re-reduction of one (post-calibration) dataset with the newer 
CASA software produced no significant differences.  As described in M01, the
angular size of the source is comparable to the size of a VLA antenna, so 
two pointings are combined to form the final image.  The dumbbell system 
NGC~326 itself has been observed with the \textit{Hubble Space Telescope} (HST)
as a snapshot observation \citep{capetti00}, and there is archival ROSAT
data (W95).  

\begin{figure*}[t]
\begin{center}
\includegraphics[width=0.9\textwidth]{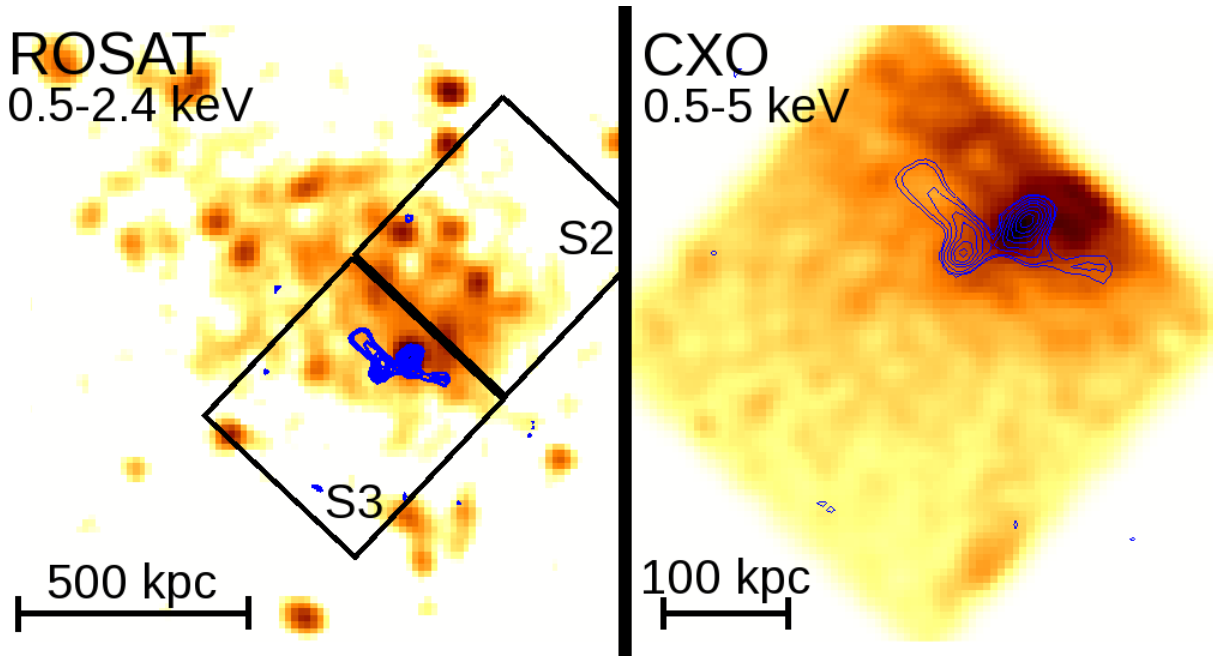}
\caption{\textit{Left:} ROSAT 0.5-2.4~keV image smoothed with a
$\sigma=3$ Gaussian kernel (no point source subtraction).  The black boxes
show the outlines of the CXO chips with cluster emission.  The radio contours
at 4.9~GHz are overlaid in blue.
\textit{Right:} CXO 0.5-5~keV image of the S3 chip with point sources subtracted and
binned to $16\times16$ pixels, then smoothed with a $\sigma=3$~Gaussian kernel.
The radio source is shown in blue as on the left.}
\label{rosat}
\end{center}
\end{figure*}

In the ROSAT PSPC image (Figure~\ref{rosat}) the X-ray peak coincides with the radio
galaxy but is not near the center of the cluster (W95).  The ROSAT image does
not have point sources subtracted; the bright spot near the southeast edge of
the CXO S3 chip is a point source subtracted in the CXO image shown in 
Figure~\ref{rosat}.  The CXO image reveals
a significant local anisotropic enhancement in the soft X-rays near this peak
elongated in the direction of the primary lobes \citep{hodges-kluck10a} that
cannot be seen in the ROSAT image.  In fact, the ROSAT X-ray peak is 
cospatial with the northern lobe of the radio galaxy and \textit{not} with the
dumbbell galaxies; these galaxies are visible in the CXO image but on scales
much smaller than the ROSAT point spread function (PSF).  

\begin{figure}[b]
\begin{center}
\includegraphics[width=0.5\textwidth]{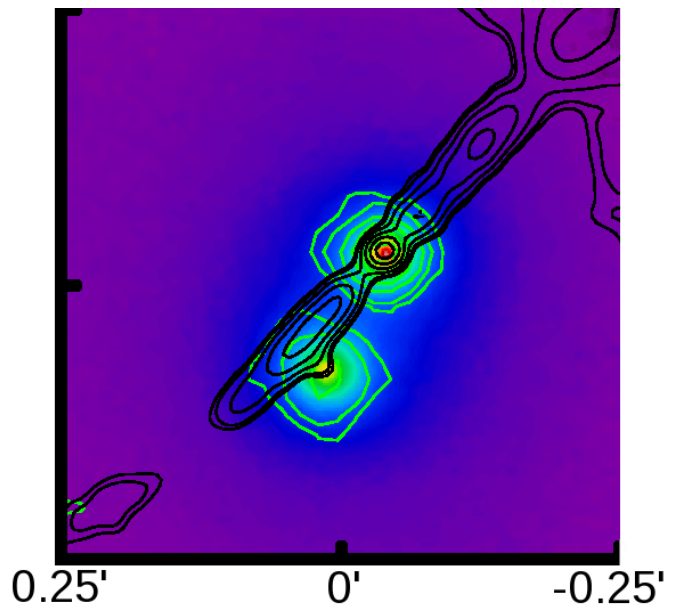}
\caption{HST WFPC2 image of NGC~326 (median-filtered to remove cosmic rays) overlaid with
X-ray contours (green) and 1.55~GHz contours (black, primarily showing the
radio jet).  The ISM of both 
components of NGC~326 are visible and resolved by CXO.  The northern component
is brighter in both the optical and X-ray, and is the host of B2~0055+26.}
\label{ism}
\end{center}
\end{figure}

\subsection{The ISM of NGC~326}

The CXO image resolves the constituents of the dumbbell NGC~326 (Figure~\ref{ism},
smoothed X-ray contours overlaid in green).
The northern component is the host galaxy of the radio source and has an obvious
strong point source at its center.  As in the HST images, the northern
component is the brighter of the pair \citep{capetti02}. 
Spectral fits for the two galaxies are described below.

\paragraph{\textit{Northern Component}}
In \citet{hodges-kluck10a} we measured a temperature of $kT \sim 0.7$~keV in this
galaxy both including ($kT = 0.68\pm0.06$~keV) and excluding ($kT = 0.7\pm0.1$~keV) the
central point source, which we fit using a power law of $\Gamma =
1.3\pm0.4$. Metallicity was frozen at solar abundances. 
This fit produces a $\chi^2 = 11.1$ for 16 degrees of freedom
(20~counts bin$^{-1}$).  
The (unabsorbed) model luminosity of the thermal component is 
$L_X \approx 7.4\times 10^{40}$~erg~s$^{-1}$, whereas the X-ray luminosity
of the power law is $L_X \approx 5.4\times 10^{40}$~erg~s$^{-1}$.  

An acceptable fit can be obtained using a 2-T fit with $kT_1 = 0.7$~keV
and $kT_2 > 3$~keV.  Presumably a higher temperature contribution
would come from the projected ICM, but this is not a satisfactory
explanation because the excess high-energy photons that require the
higher temperature or power law component are concentrated inside the
PSF at the location of the radio point source.  Meanwhile, photons
outside the half-power distance of the PSF are preferentially
\textit{softer}, suggesting that they are really part of the extended
ISM.  An AGN origin for the harder X-ray emission is therefore favored.

\begin{figure*}[t]
\begin{center}
\includegraphics[width=0.48\textwidth]{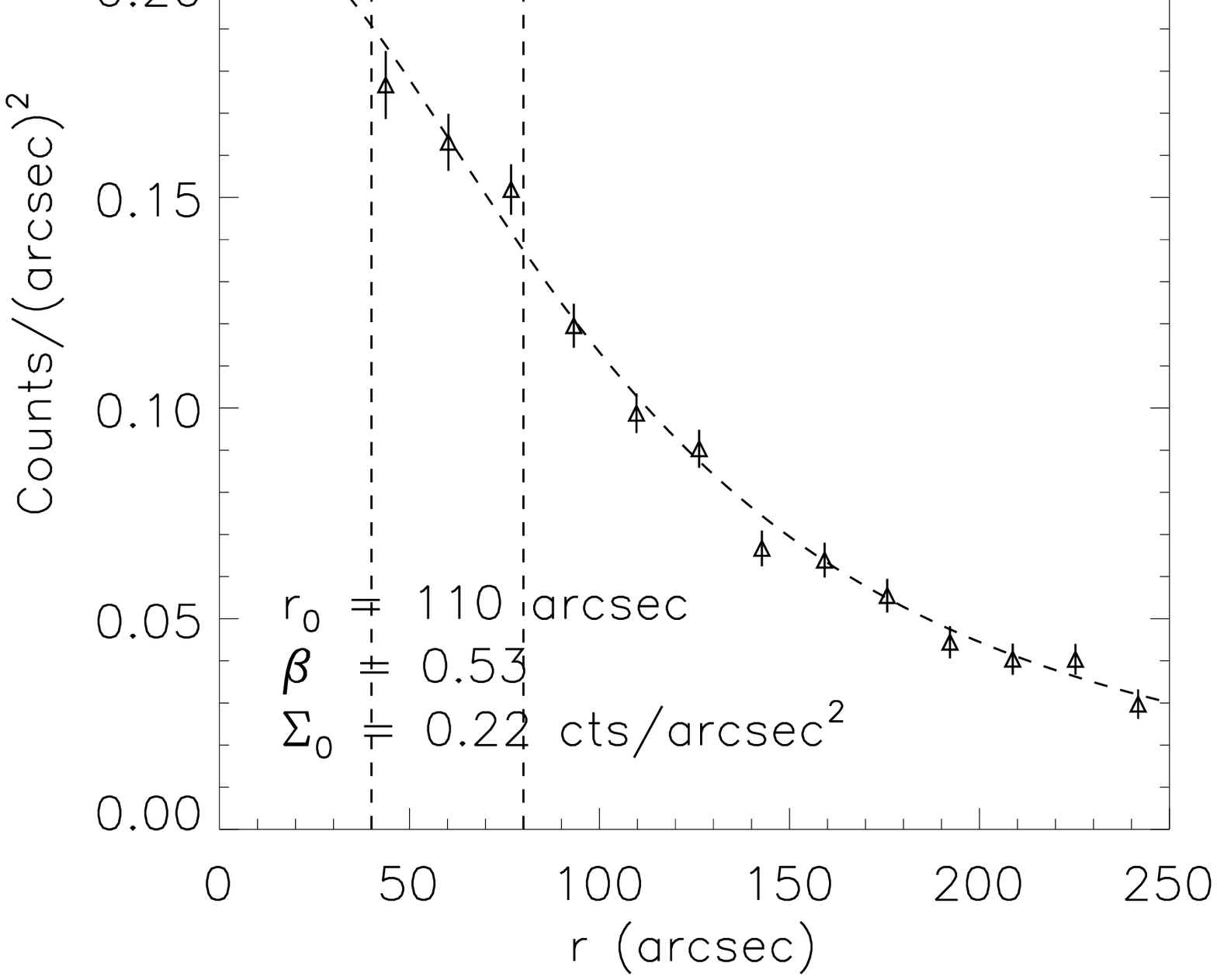}
\includegraphics[width=0.48\textwidth]{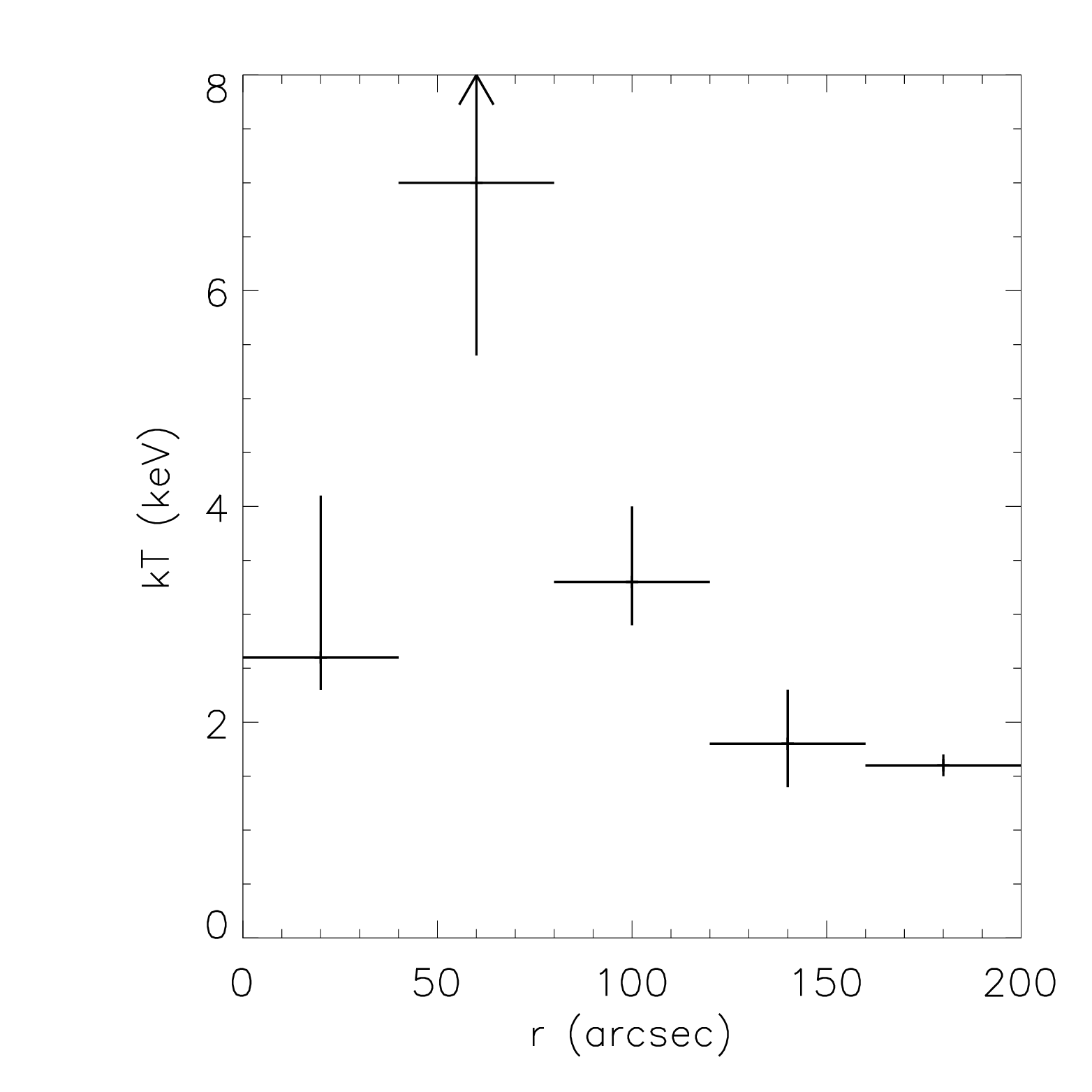}
\caption{\textit{Left:} 1D radial profile extracted from the ACIS-S3 chip centered on NGC~326.
Point sources have been extracted as well as extended ISM emission.  The
best-fit 1D $\beta$ model is overplotted.  Dashed lines denote the region covered by the high
temperature bin in the right panel.
\textit{Right:} Deprojected temperature profile for annuli 40$^{\prime\prime}$ 
in width, centered on NGC~326.  Temperatures and errors obtained using 
{\tt steppar} with {\tt projct(phabs*apec)}.  The temperature
in the bin associated with the high-temperature front is not well constrained,
but must be above $\sim 5.5$~keV.
}
\label{radial_profile}
\end{center}
\end{figure*}

\paragraph{\textit{Southern Component}}
The southern component is fainter and has no obvious nuclear
point source.  Only 154 counts are detected, with a total luminosity
of $L_X \sim 6\times 10^{40}$~erg~s$^{-1}$. Despite the small number
of counts, the X-ray spectrum is poorly fit by a single 1-T model
($\chi^2 \sim 12$~for 8 degrees of freedom) but a good fit is achieved
by including a power law component with a frozen photon index $\Gamma
= 2$ ($\chi^2 = 7.3$ for 8 degrees of freedom). There are insufficient
counts to constrain $\Gamma$, but a power law interpretation is
preferred as in the northern component since the harder photons are
concentrated toward the center of the detect cell.  Some power law
emission is expected from unresolved X-ray binaries 
\citep[see, e.g.][]{diehl07}, but we cannot rule out a weak nuclear
source due to the 1~GHz emission from the nucleus.

A common envelope of diffuse optical emission surrounds the components
of the dumbbell galaxy, but no similar envelope is detected in the
X-rays.  The radio image leaves open the possibility of interaction
between the jet and the ISM of the southern component, but no X-ray
jet is detected and the signal-to-noise ratio ($S/N$) in the southern
component is too low to search for disturbances.  

\subsection{Cluster Atmosphere}

The cluster X-ray emission is diffuse and extensive, covering the ACIS-S2 and 
S3 chips as well as some of the I2 and I3 chips.  No other chips were active
for this exposure.  The aimpoint for the observation was on the S3
chip, so it is difficult to identify and excise point sources on the I2 and I3
chips.  While bright point sources are
still detected with algorithms such as {\tt wavdetect}, faint point sources
are distributed over many pixels and are difficult to distinguish from the
cluster atmosphere or the background.  Thus, we focus our efforts on the S2 and
S3 chips (the radio galaxy is wholly contained in the latter).  Quantities
are listed in Table~\ref{clusterprops}.

\begin{figure*}[t]
\begin{center}
\includegraphics[width=0.9\textwidth]{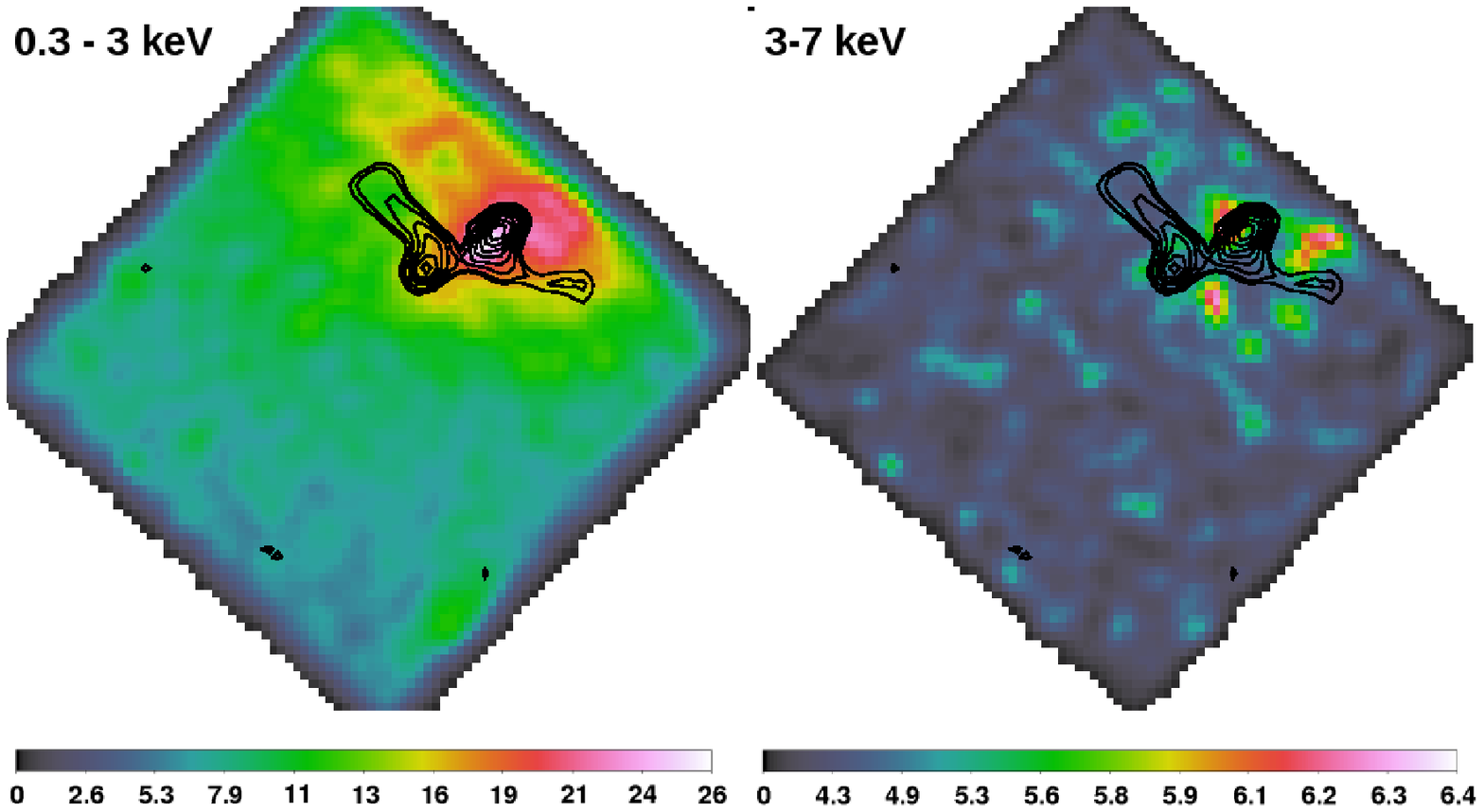}
\caption{\textit{Left}: 0.3-3~keV CXO image of S3 chip, binned to $16\times16$ pixels
then smoothed with a $\sigma=3$~px Gaussian kernel.  4.9~GHz contours are overplotted.
The units of the scale are in counts and the colormap is linear.
\textit{Right}: 3-7~keV image processed in the same way as left except that the
units of the scale are exponential to highlight the brightest regions in the
map.}
\label{hardness_compare}
\end{center}
\end{figure*}

\paragraph{\textit{Large-scale structure}}
We generated a radial profile from the point-source subtracted images, centered
on the X-ray peak near NGC~326 (Figure~\ref{radial_profile}), subtracting
local background from an annulus outside the fit region.  We fit a 
1D $\beta$ model (a Lorentzian of form $f(r) = A_0 [1+(r/r_0)^2]^{(-3\beta+1/2)}$,
where $A_0$ is the peak amplitude in counts per square arcsecond and $r_0$ is the
core radius in arcseconds) and find a good fit for $\beta = 0.53 \pm 0.04$
and $r_0 = 110 \pm 10$~arcsec (100~kpc).  This gives a $\chi^2 = 11.10$ 
for 11~degrees of freedom (using CIAO's \textit{Sherpa} $\chi^2$ with the
Gehrels variance function).  Our value of $\beta$ is consistent with the W95
value, but they find a core radius $r_0 \sim 180$~kpc in WMAP cosmology.
This may be due to the larger field of view of the ROSAT
image and the fact that our background annulus encompasses some
cluster emission.  There is also obvious anisotropy in the cluster on
the scale of the S3 chip (Figure~\ref{rosat}), so our fit is only a
good description of the atmosphere in the vicinity of NGC~326.  Since
the X-ray emission may be a superposition of multiple clusters or
groups (W95), we may be isolating
group-scale gas associated with NGC~326. 

\begin{figure}
\begin{center}
\includegraphics[width=1.0\linewidth]{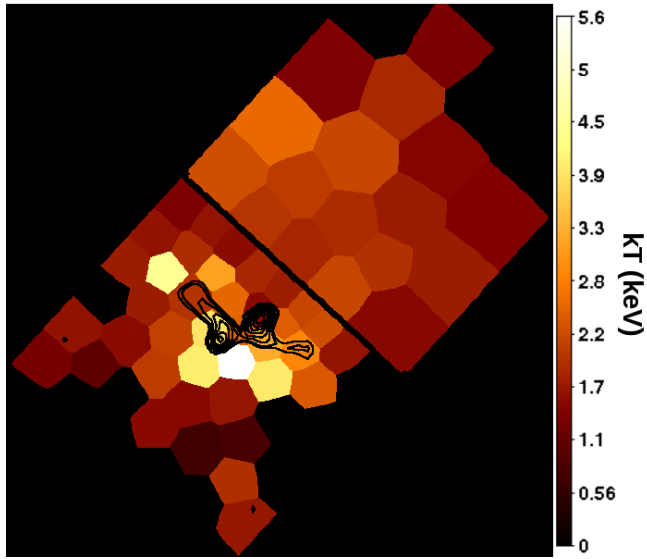}
\caption{Temperature map made from adaptive bins with $S/N = 35$ each.  The
bins on the S2 chip are larger owing to the lower surface brightness.  
Temperatures are measured from 1-T {\tt apec} fits; blacked out bins represent
bins for which a temperature could not be derived with an acceptable fit.  4.9~GHz
contours of NGC~326 are overlaid for reference.}
\label{tempmap}
\end{center}
\end{figure}

W95 measured a cluster temperature of approximately 2~keV and a metallicity 
of 0.3~times solar from the ROSAT data.  The CXO data is largely in agreement,
with temperatures ranging from $kT \gtrsim 2$~keV near
the core to $kT \sim 1.5$~keV near the edge of the cluster.  However, a temperature
map made using weighted Voronoi tessellations to adaptively bin the images into
bins with $S/N = 35$
\citep{diehl06,cappellari03} shows a ``front'' of high-temperature plasma to
the southeast of the radio galaxy (Figure~\ref{tempmap}).  The
temperatures in bins in 
this front range from $3-6$~keV for a fixed metallicity of $Z =
0.3Z_{\odot}$ (W95).  The temperature is not well
constrained, but no acceptable fits were found below $kT \sim 2.5$~keV
for different bin sizes, metallicities, or 2-T fits.  

Another way to estimate temperatures in the front is to ``deproject'' the
emission by assuming a spherical atmosphere (within 100~kpc of
NGC~326) and using annuli to break the emission into multiple bins.
In this assumed geometry, a line of sight through the center
sees a superposition of all the cluster emission, whereas a line of
sight near the edge sees only the emission from the outskirts.  Thus,
each annulus is corrected for the annuli exterior to it, hopefully
mitigating the effects of projection.  Since the high-temperature
front is about 70~arcsec from the core and the core radius is $r_0 =
110$~arcsec, we would expect significant contamination and that the
deprojected temperature would be higher.  This is indeed what we find
(Figure~\ref{radial_profile}), but we lack the counts to decisively
bound the temperature.  Since a power law model is not a good fit, the
detection of hot plasma southeast of the radio source seems secure. 
We discuss the nature of this front in Section~3.  

In addition to the high-temperature front and surface brightness enhancement
around the radio galaxy, the CXO map reveals a few small-scale
structures unresolved by ROSAT. 

\paragraph{\textit{High-temperature knots}}
First, 
separating the data into soft ($0.3-3$~keV) and hard ($3-7$)~keV bands reveals
the presence of bright knots around the radio galaxy in the hard X-ray band 
(Figure~\ref{hardness_compare}).   Above 7~keV, the image is very noisy.  
The presence of bright knots around the radio emission suggests 
interaction between the radio source and the ICM.  This inference of association is 
strengthened by the observation that the western wing appears to pass through
a narrow channel devoid of hard X-ray emission in Figure~\ref{hardness_compare}.
Although higher signal is required to establish their morphology, the data
suggest that these knots are part of high-temperature rims bounding the 
radio source: the hard X-ray emission appears to avoid the radio emission.
The presence of the brighter knots is consistent with the detection of X-ray
emitting rims around radio sources \citep[][and references therein]{mcnamara07}.
However, usually such rims are cool, whereas these features are distinct only
in the hard $3-7$~keV band.

There are $\sim$500 total events in the $3-7$~keV band in these
features.  When spectra are extracted from these regions, the spectra
are fit well by 1-T thermal models with 
$kT \sim 3.2$~keV.  Given that we view the cluster in projection and the surrounding
material has $kT \sim 2$~keV, this is a lower
bound on the temperature in the knots.  Despite the small number of
counts, a nonthermal origin for the hard emission is disfavored.  If
we fit the spectrum with a thermal model frozen at the $kT \sim 2$~keV
temperature of the background plasma and add a power law component, we
require $\Gamma = 1.3^{+0.2}_{-0.1}$ for a good fit, corresponding to
a spectral index $\alpha = 0.3$.  There are two plausible
possibilities for nonthermal emission resulting from interaction of
the ICM with the edge of a radio galaxy: inverse Compton emission and Fermi
acceleration from shocks.  In the latter case, we would expect $\alpha
= 0.7$, so we can rule this out.  Inverse Compton emission is also
unlikely in this case.  The spectral index of the radio galaxy is
$\alpha \sim 0.7$ from 25~MHz through 1~GHz, so if it is synchrotron
self-Compton, the boosted electrons would have to come from a
population below 25~MHz.  The average frequency boost is
$\nu_{\text{obs}} \sim \tfrac{4}{3}\nu_0 \gamma^2$, so this would
imply $\gamma \gtrsim 2\times 10^5$!  Moreover, the knots are outside
the radio emission at GHz frequencies, and we would expect to see GHz
emission from these regions as well.  Supposing that the inverse
Compton emission comes from a different source, we would still expect
to see synchrotron emission in the GHz band from the boosted
electrons if $B_{\text{knot}} \approx B_{\text{eq}}$ (discussion
follows in Section~3).  Thus, we conclude that the knot emission is
thermal. 

\begin{figure}
\begin{center}
\includegraphics[width=0.5\textwidth]{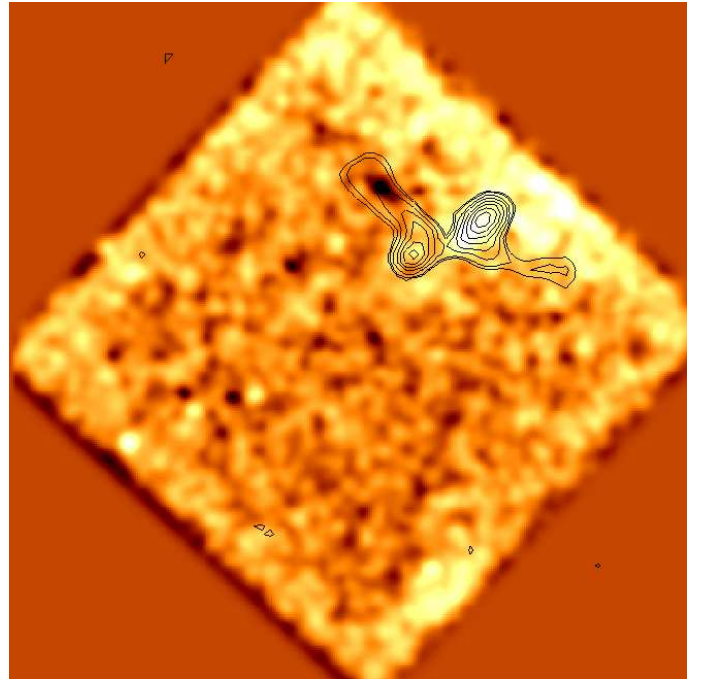}
\caption{Unsharp mask image made using a radius of 50~pixels, a 5\% contrast 
increase, and a threshold of 0 counts.  Zero is represented by the uniform
orange color outside the chip.  The image is binned to $16\times16$ 
pixels and smoothed by a Gaussian
kernel of $\sigma=3$~px.  The enhancement at the chip edges is
not physical, and the significance of the decrement associated with the eastern
wing is based on an unsharp mask on only those pixels making up chip S3.  The
negative associated with the wing is by far the deepest in the image.}
\label{cavity}
\end{center}
\end{figure}

\paragraph{\textit{Wing Cavity}}
In addition to positive features, an unsharp mask image of the S3 chip
(Figure~\ref{cavity}) reveals an X-ray decrement associated with the east
wing of NGC~326 centered about 95--10~kpc from the host.  
Although there are other X-ray decrements in the field,
this is the deepest.  
The unsharp mask uses a radius of 50~pixels, a contrast enhancement of 5\%, and
a threshold of 0~counts to apply the unsharp mask.  For viewing ease 
we show the entire chip smoothed by a Gaussian kernel with
width $\sigma = 3$~pixels, but the chip edges enhanced by the unsharp mask are
unphysical and should be ignored.  This decrement is also plainly visible in
a residual image produced by subtracting off the $\beta$-model fit above (not
shown).  The decrement is in a shallow trough in the exposure map,
but the trough spans the length of the chip and our exposure corrected image
demonstrates that the decrement described here is substantially deeper than
the trough.  We tentatively
identify this decrement as a ``cavity'' associated with the eastern wing.

Computing the significance of the decrement is not 
straightforward because of the obvious anisotropy of the cluster emission on 
scales larger than the chip.  Subtracting off the $\beta$-model reveals a
gradient in the residuals apart from the decrement.  There is no obvious way
to account for this emission, since increasing the number of \textit{ad hoc} 
model components risks overconstraining the problem.  However, we can at least
ask whether a cavity at the location in the cluster could be detected by
using the expressions in \citet{diehl08}.  Equation~13 in their paper gives
the $S/N$ of a cavity of a given size and at a given distance from the cluster
core as a function of $\beta$, $r_0$, $kT$, $n_0$, and observational
parameters such as the detector effective area and exposure time.  For our
cluster ($kT \sim 2$~keV, $n_0 \sim 3\times 10^{-3}$~cm$^{-3}$, $\beta \sim 0.5$,
$r_0 \sim 100$~kpc), a cavity at the position of the decrement 
would require a radius of $r_{\text{cav}} \sim 21$~kpc in order to 
be detected with a $S/N \gtrsim 3.0$ in our CXO exposure.  

This radius is consistent with the X-ray decrement seen in Figure~\ref{cavity}
and the width of the eastern wing on the sky; a real cavity is unlikely to be
larger than this.  Hence, the decrement seen in the unsharp mask and residual
image is consistent with the detection of a cavity. 
A deeper exposure is required to firmly establish its presence:
a higher effective area (i.e., XMM-\textit{Newton}) or a 
$t_{\text{obs}} \sim 400$~ks exposure with CXO would easily detect a
20~kpc cavity at more than 7$\sigma$. 

\begin{deluxetable}{lll}[t]
\tablenum{1} 
\tabletypesize{\scriptsize} 
\tablecaption{Derived Parameters} 
\tablewidth{0pt} 
\tablehead{ 
\colhead{Parameter} & \colhead{Value} & \colhead{Units}
}
\startdata
\multicolumn{3}{c}{Cluster Parameters}\\
\cline{1-3}\\
$r_0$      & 110$\pm$5         & arcsec \\
$\beta$    & 0.53$\pm$0.04     &        \\ 
$A_0$      & 0.22$\pm$0.02     & cts/(arcsec)$^2$\\
$kT$       & $\sim$2           & keV (avg)\\
$n_0$      & $3\times10^{-3}$  & cm$^{-3}$\\
$P_0$      & $\sim10^{-11}$    & dyne cm$^{-2}$\\
$c_s$      & 500               & km s$^{-1}$\\
\cline{1-3}\\
\multicolumn{3}{c}{Radio Galaxy}\\
\cline{1-3}\\
N. Lobe Length & $\sim$50      & kpc\\
S. Lobe Length & $\sim$75      & kpc\\
E. Wing Length & $\sim$100     & kpc\\
W. Wing Length & $\sim$100     & kpc\\
1 GHz $t_{\text{sync}}$ & 70   & Myr\\
$B_{\text{eq}}$ & $\sim$7      & $\mu$G\\
\enddata
\tablecomments{\label{clusterprops}
Measured and derived properties of NGC~326 and environs.  Where errors are quoted, they are
90\% confidence limits.  Otherwise, quantities should be treated as approximate with 
justification in text.  The radio structures are seen in projection and may be longer than
the given values.}
\end{deluxetable}

\section{Derived Properties}

Properties derived in this section are summarized in Table~\ref{clusterprops}.

\paragraph{\textit{Sound Speed and Terminal Buoyant Velocity}}
Whereas powerful radio galaxies grow hypersonically as their jets plough through
the tenuous IGM, the low-power FR~I sources are thought to expand transonically
or slower over the bulk of their lifetime \citep[for a typical model, see][]{luo10}.  
The remnants of dead radio galaxies are also thought to rise buoyantly in
the ambient medium, and the growth of the wings of XRGs/ZRGs may be subsonic
depending on the mechanism.  Hence, the sound speed is a useful yardstick to
apply to these sources. 

Within 100~kpc of the cluster core, the average temperature in regions other
than the high-temperature front is about $kT \approx 2$~keV, yielding a sound 
speed
\begin{equation}
c_s = \sqrt{\frac{\gamma kT}{\mu}} \lesssim 500 \text{ km s$^{-1}$} = 0.5 \text{ kpc Myr$^{-1}$}, 
\end{equation}
in the approximation of an isothermal relaxed atmosphere and $\mu =
1.4m_{H}$.  $\gamma$ is taken to be 5/3.

The terminal buoyant velocity of a light spherical bubble suspended in a medium
is given by
\begin{equation}
v_{\text{buoy}} \approx \sqrt{\frac{2gV_b}{C_D A_b}} \sim c_s \sqrt{\frac{16}{3\gamma}\frac{r_b}{r_0^2}\frac{r}{1+(\tfrac{r}{r_0})^2}},
\end{equation}
where $V_b$, $A_b$, and $r_b$ are the volume, cross-sectional area, and radius of
the bubble respectively, $g$ is the local gravitational acceleration, $C_D \approx 0.75$ is
the drag coefficient, and $r_0 \approx 100$~kpc is the core radius of the
atmosphere.  For the cavity at the edge of the eastern wing, $r \sim 100$~kpc
and $r_b \sim 20$~kpc, so 
\begin{equation}
v_{\text{buoy}} \sim 0.8c_s = 0.4 \text{ kpc Myr$^{-1}$}.  
\end{equation}

\paragraph{\textit{Equipartition estimate of $B$}}
Following \citet{tavecchio06}, we estimate the equipartition field $B_{\text{eq}}$
from the 1.4 and 4.9~GHz maps of the active lobes:
\begin{equation}
B_{\text{eq}} \delta = \biggl(\frac{2\times 10^{-5} \gamma_{\text{min}}^{1-2\alpha} \nu_s^{\alpha} L_s(\nu_s)}{(2\alpha-1) c(\alpha) V}\biggr)^{1/(3+\alpha)}
\end{equation}
where $\delta$ is the Doppler factor, 
$\gamma_{\text{min}}$ is the minimum electron cutoff energy, $\nu_s$ is
the synchrotron frequency (we assume the observed frequency $\nu_0 = \nu_s$),
$L_s$ is the synchrotron luminosity in volume $V$, and $c(\alpha)$ is a
numerical factor in \citet{pacholczyk70}.  The
lobe volume is approximately $V \sim \pi r^2 h$, where $r$ and $h$ are the 
measured half-width and length of the lobes respectively: $r \sim 20$~kpc and
$h \sim 75$~kpc (south) or 50~kpc (north).  Hence, $V \sim 5\times 10^{69}$~cm$^{-3}$.
The luminosity is
\begin{equation}
L_s (\nu_s) = 4\pi D_L^2 F_s(1+z)^{\alpha-1},
\end{equation}
where $D_L = 210.3$~Mpc is the luminosity distance.  High quality images exist at
multiple frequencies, showing no evidence for spectral steepening between 1.4 and 4.9~GHz
($\alpha_{1.4-4.9} \sim 0.7$; M01).  Since $B_{\text{eq}}$ depends on the quantity
$\nu_s^{\alpha} L_s (\nu_s)$, estimates of $B_{\text{eq}}$ at 1.4 and 4.9~GHz must
come out to be the same. 

The flux at 1.4~GHz is $F_{s,1.4} \approx 0.86$~Jy, while at 4.9~GHz
$F_{s,4.9} \approx 0.42$~Jy.  These values correspond to 
$B_{\text{eq}} \delta \sim 7 \mu$G at 1.4~GHz and $\sim$6~$\mu$G at 4.9~GHz.  Since
the luminosity in these images is dominated by lobe emission, presumably
$\delta \sim 1$.  Hence, we adopt
\begin{equation}
B_{\text{eq}} = 7 \mu\text{G}
\end{equation}
as the equipartition field in the lobes. 

From the synchrotron frequency, $\nu_s = 4\times 10^6 B \gamma^2$~Hz, we then
find $\gamma \sim 7100$, yielding a synchrotron cooling time of 
\begin{equation}
t_{\text{sync}} \sim 2400 \gamma_{4}^{-1} B_{-6}^{-2} \text{ Myr} \sim 70 \text{ Myr}
\end{equation}
for the 1.4~GHz electrons, where $\gamma_{4}$ is $\gamma$ in units of $10^4$ and
$B_{-6}$ is $B$ in units of $\mu$G.  For the 4.9~GHz electrons, 
$t_{\text{sync}} \sim 37$~Myr.  For our purposes,
the synchrotron cooling time is more relevant in the wings, which will fade from
view on this timescale without a fresh injection of electrons.  
Since the wings are bright at 4.9~GHz, 
 the most recent injection must be no more than about 40~Myr old.  

\paragraph{\textit{High-temperature Front/Bow Shock}}

The high-temperature front visible in the temperature map (Figure~\ref{tempmap})
and evident in the deprojected temperature profile (Figure~\ref{radial_profile})
follows the contours of the soft X-ray emission to the southeast, and is 
slightly within the core radius $r_0$ of the 1D surface brightness profile.

The temperature of this feature is uncertain due to the low
number of counts.  The temperature from the adaptive binning scheme is
between $kT \sim 3.5-5$~keV, whereas the deprojected temperature must
be $kT \gtrsim 5.5$~keV, at least 
$2-3$~times greater than the surrounding medium.  We now consider
whether this feature is consistent with a shock.

The Rankine-Hugoniot shock-jump conditions are
\begin{eqnarray}
\frac{\rho_2}{\rho_1} & = & \frac{v_1}{v_2} = \frac{(\gamma+1)M_1^2}{(\gamma-1)M_1^2 + 2}\\
\frac{P_1}{P_2}       & = & \frac{\rho_2 k T_2/m}{\rho_1 k T_1/m} = \frac{2\gamma M_1^2 - (\gamma-1)}{\gamma+1}\\
\frac{T_2}{T_1}       & = & \frac{[(\gamma-1)M_1^2 + 2][2\gamma M_1^2 - (\gamma-1)]}{(\gamma+1)^2 M_1^2},
\end{eqnarray}
where $\gamma = 5/3$ is the adiabatic constant and $M_1 \equiv v/c_s$ 
is the Mach number.  If we take $T_2/T_1 \approx 2$, then $M_1 \approx 1.9$. 
Incorporating uncertainties from the thermal models and binning scheme, 
we find 
\begin{equation}
M_1 \sim 1.8 - 2.7.
\end{equation}
Assuming $M_1 \sim 2$, we would expect a
downstream density jump of $\rho_2 = 2.3\rho_1$, with a corresponding
downstream surface brightness $\sim$5~times higher.  However, farther
downstream compressed material will relax, and the shock front is thin
relative to the atmosphere.  Thus, while we would expect to see a
surface brightness discontinuity, it might be substantially smaller. 

In fact, we do not see such a discontinuity.  The 1D radial profile
shows a hint of a surface brightness jump with $\Sigma_1/\Sigma_2 \sim
1.4$ in a region nearly cospatial with the temperature jump
(Figure~\ref{radial_profile}), but there are insufficient counts to
conclusively identify a discontinuity or detect the expected lag
between the discontinuity and rise in temperature.  Another hint is
the elliptical surface brightness enhancement co-aligned with the jet
axis (Figure~\ref{hardness_compare}).  This enhancement was earlier
interpreted by us as the pre-outburst
structure of the ICM, but it may result from the outburst.
Unlike cold fronts in which a surface brightness discontinuity occurs
in pressure equilibrium, the X-ray bright gas in the elliptical region 
has a somewhat higher pressure than outside.  A deeper observation is
required to confirm the discontinuity.

Another objection to the shock hypothesis is that the front is
observed only on one side, but this can be explained by the fact that
the cluster atmosphere brightens to the northwest and tapers off to
the southeast (Figure~\ref{rosat}).  Shock-compressed gas will be most
visible against a dim background, and the high-temperature knots near
the northwestern lobe suggest that there is indeed shocked gas in the
vicinity of the radio galaxy on the northern side.  If the cluster
atmosphere is denser in this direction, the shock wave will sweep up
more material and advance more slowly per unit time than the
southeastern front.  The northwestern lobe \textit{is} curtailed 
relative to the southeastern one,
but $\sim$4~times more photons are required to measure a density
gradient on the relevant scales.

If the high-temperature front is not a shock from the radio galaxy,
its origin is not obvious.  The radiative (free-free) cooling time of the hot gas
in the region is more than 2~Gyr, but conduction due to hot electrons
would also tend to spread out the thermal energy.  It is possible this
feature is a shock associated with a prior outburst that has since swept
up material and become a large sound wave, but if it is traveling at the
sound speed its distance from the active lobes would imply but a short delay 
between successive outbursts.   It seems less likely that
magnetic structures could produce a region (via anisotropic conduction)
so much hotter than the surroundings.  Without better data, we tentatively
conclude that the feature is best explained as a shock associated with the
current outburst. 

\paragraph{\textit{Active Lobe Age and Jet Power}}

Assuming that the active lobes represent a single outburst and inflate bubbles
as large as the region of radio emission, we can make a rough estimate of the
jet power from the size of the active lobes using standard arguments: 
the average jet power is the cavity energy divided by the age of the outburst.

A note of caution is required because no cavity has actually been detected
associated with the active lobes.  Although it seems clear that radio jets
must form bubbles in the ICM even when no cavity is detected, the filling factor
of older cavities by ambient material is unknown.  Assuming cylindrical 
cavities the volume depends only linearly on the projected length, so 
we sketch out order-of-magnitude estimates to place
limits on the jet power.  

The energy required to blow a bubble in the ICM is $E_{\text{cav}} = \gamma pV/(\gamma-1) = 4pV$,
where $p$ is the pressure, $V$ the volume, and $\gamma = 4/3$ the adiabatic
constant for relativistic electrons.  The cavity enthalpy can be
expressed as
\begin{equation}
E_{\text{cav}} = 4pV \sim 4\pi r^2_{\text{lobe}} h_{\text{lobe}} n_{\text{gas}} kT_{\text{gas}}
\end{equation}
where the gas subscript refers to the ICM.  $r_{\text{lobe}} \sim 20$~kpc and 
$h_{\text{lobe}} \sim 75$~kpc are estimates based on visual inspection of the
M01 maps and subject to projection and resolution uncertainties;
lower frequency maps suggest that a cylinder is a good representation.  
$kT_{\text{gas}} \sim 2.5$~keV
is obtained from 1-T fits to the region, and $n_{\text{gas}}$ is likewise 
obtained from the spectral fit via the model emission measure.  For assumed
geometry we follow the method in \citet{mahdavi05}, in which the emission
is assumed to come from a spherical shell with outer radius $R_1$ and inner
radius $R_2$.  The line-of-sight depth is then $2(R_1 - R_2)^{1/2}$. 
Near the active lobes, 
$p \sim 10^{-11}$~dyne~cm$^{-2}$.  Hence, $E_{\text{cav}} \sim 10^{59}$~erg.
Note that $h_{\text{lobe}}$ is less certain than $r_{\text{lobe}}$ because of projection effects.
Hence, if the wings were formed in the same outburst, the energy is only
greater by a factor of $\sim$2.

A plausible range for the jet power can be obtained in view of the
putative shock front.  The presence of a shock suggests supersonic expansion,
at least in the past---the radio jet is now more akin to that of an FR~I
source where subsonic expansion is expected.  The Mach~$\sim$2 shock implies
a source age of less than 75~Myr, but since shocks sweep up
material and slow down, the initial expansion was likely substantially
faster.  The well defined lobes that still give the radio galaxy an
edge-brightened morphology suggest that the transition occurred 
recently, which is consistent with estimates of FR~II and quasar lifetimes of tens of Myr
\citep{bird08,martini04}.  
The shock front is also still relatively close to the
radio galaxy, so a scenario in which the jets are simply illuminating a very
old source is disfavored. 

An average jet power of 
$P_{\text{jet}} \sim 10^{44}$~erg~s$^{-1}$, corresponding to a lobe age of
32~Myr, is consistent with the data.  
For reference, if NGC~326 had jet power commensurate with Cygnus~A
\citep[$P_{\text{jet}} \sim 10^{46}$~erg~s$^{-1}$;][]{stockton94},
its lobes would be inflated in just 0.3~Myr!  There is no obvious cocoon and
the source lacks hot spots, so this is clearly too high.  On the lower end,
transonic lobe expansion gives a jet power of $P_{\text{jet}} \sim 5\times 10^{43}$~erg~s$^{-1}$,
which is inconsistent with the shock front.  $P_{\text{jet}} \sim 10^{44}$~erg~s$^{-1}$
is therefore probably accurate to within a factor of $2-3$, giving an age 
between 10--50~Myr.  
This power is akin to Perseus~A \citep{fabian06}.

\section{Discussion}

\subsection{Reconstructing the History of NGC~326}

In view of the high resolution X-ray data, we re-examine the formation history
of NGC~326 and its long wings.  The salient results of the X-ray analysis 
follow: (i) The very long wings in a relatively cool environment require a 
single outburst to have an unrealistic age if they expanded transonically or 
slower.  (ii) The putative shock and high-temperature rims appear to be 
causally connected to the radio outburst, but are inconsistent with the 
present-day jet, which strongly resembles an FR~I source: the jet decelerates
and breaks up before reaching the ends of the lobes, and no terminal hot spots
are observed.  Thus, we do not expect that the strong backflows associated with
the hot spots of FR~II sources are present in NGC~326.  Still, it is
notable that the active lobes are well defined and bright, which led to the
original classification of the source as an FR~II.  (iii) The  
radio galaxy is situated at the region of brightest X-ray emission in the 
cluster, but the cluster emission shows structure on scales slightly smaller 
than the chip.  The wings appear to follow an edge in this emission
(Figure~\ref{rosat}).  

The data suggest that the outburst that produced the radio galaxy was much
more powerful in the past.  NGC~326 is very weak for an FR~II source relative
to its host galaxy luminosity \citep{cheung09}; if the present-day kinetic
luminosity of the jet is in equipartition with the radiative luminosity, the
jet power is on the order of $P_{\text{jet}} \sim 10^{41}$~erg~s$^{-1}$.  
Although equipartition need not be the case, the jet would need to be extremely radiatively
inefficient in order to drive the shock and inflate the cavities if it is.  The
brightness of the jet near the core suggests this is not the case.  If
NGC~326 is the result of a powerful outburst that decayed, the shock and
well defined lobes are natural consequences of the more powerful stage.  

In fact, the parameters of such an outburst fall within the accepted parameters
for radio-galaxy power and lifetime as well as the energy required in this
system.  Suppose that the jet power was initially 
$P_{\text{jet}} \sim 10^{45}$~erg~s$^{-1}$, more powerful than Perseus~A but
less powerful than Cygnus~A, and decayed exponentially on an $e$-folding time 
of 10~Myr \citep[thought to be a typical FR~II lifetime;][]{bird08}.  Then the total
energy input by the jet into the ICM is about $3\times 10^{59}$~erg, of which
about two-thirds is injected during the FR~II phase ($0-10$~Myr).  
After an e-folding time, the power of this hypothetical
jet is only a few times higher than the observed radio power of the core, and
the energy injected into the ICM is within a factor of 2 of our cavity enthalpy
estimate, shock energy and cosmic ray heating notwithstanding.  Hence, a 
fast-rise exponential-decay scenario seems to fit NGC~326.  

We can go a step farther since we know from the synchrotron cooling time that
the plasma radiating at 5~GHz in the wings cannot be older than about 40~Myr,
and from spectral aging arguments \citep[following][using the
  spectral data of M01]{myers85} that even the plasma radiating at 1.4~GHz
is only about 10~Myr old.  If this
plasma was deposited by flows from the primary lobes to the wings, the wing
length of $\gtrsim$100~kpc implies flow speeds of $3-10\times 10^{3}$~km~s$^{-1}$,
suggesting strong backflows (as in W95).  Such backflows are associated with
the hot spots of FR~II sources \citep[e.g.][]{leahy84,antonuccio-delogu10}.
The radio galaxy must then have been strong enough to form hot spots no more
than a few tens of Myr ago.  Note that this argument does not rely on a specific
mechanism of wing formation as long as the wings are illuminated by active backflows.
Such an age is consistent with the  location and tentative strength of the 
shock, which imply an outburst age of (significantly) less than 75~Myr.  
If so, the length of the active lobes implies supersonic
expansion, which is what we would expect from a jet driving a shock.

But where did the wings come from?  There are two general
ideas in the literature.  First, it is possible that the wings are
radio lobes which happen to grow in a different direction from the
jet axis due to anisotropic pressure gradients in the hot
atmospheres \citep[e.g. W95,][]{leahy84,capetti02}.  The other
possibility is that the wings represent fossil lobes from a prior
outburst in a different direction
\citep[e.g.][]{dennett02,merritt02}.  As NGC~326 has been used
as the poster-child of several ideas presented in the literature, we
will evaluate both hydrodynamic and reorientation models.  

Powerful radio galaxies generate strong backflows at the hot spot
zones of the jet head.  These flows are quite rapid \citep[up to 20\%
  of the jet speed;][]{antonuccio-delogu10} despite the relatively
slow advance of the jet head itself.  W95 argue that these flows are
buoyant in the ICM and, though rapid, may be deflected by a region of
higher pressure.  They find that flows of $\sim$3000~km~s$^{-1}$ would
be buoyant in the ICM of NGC~326 and propose that they formed the
wings.  However, it is not likely that wings formed in this way
advance at a speed close to the backflow speed.  Simulations indicate
that lobes are turbulent and that rapid backflows experience shocks
within the lobe itself, dissipating bulk kinetic energy \citep[e.g.][]{hodges-kluck11}.
Although magnetic fields may collimate flows, mature
magnetized lobes still become quite turbulent \citep{huarte-espinosa11}.  In our
recent work specifically investigating the action of pressure
gradients on radio lobes in pure hydrodynamic simulations, we found
that wings expand at most transonically due to this dissipation of
backflow thrust.  Hence, the presence of strong
backflows does not guarantee rapid wing expansion (indeed, the
internal sound speed of the lobes may be comparable to the backflow
speed).  In fact, it would be surprising if the simulations showed
otherwise, since this would lead to wings significantly wider than the
active lobes that also propagate faster than the jet head.  

\begin{figure}
\begin{center}
\includegraphics[width=0.5\textwidth,angle=90]{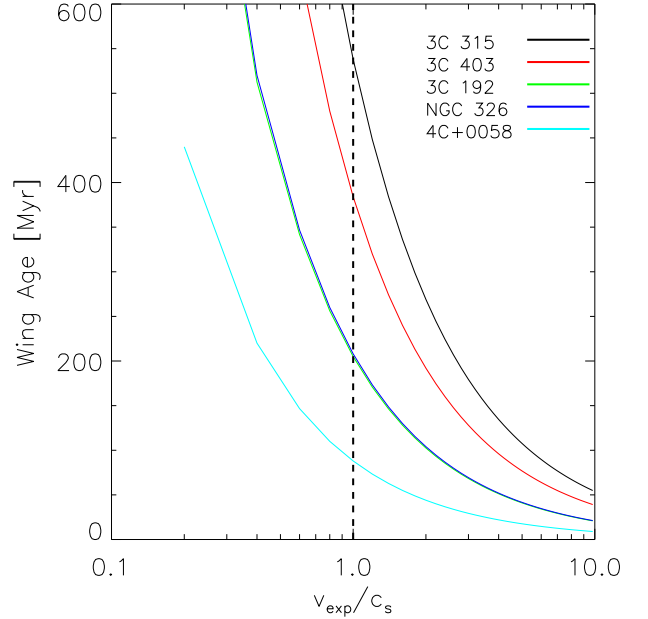}
\caption{Wing expansion times as a function of the Mach number for several XRGs
  with X-ray data.  Note that because galaxy group atmospheres are
  relatively cool, modest expansion rates may still be very
  supersonic.  Wing formation models that grow wings at a fraction or
  multiple of the sound speed require old wings.}
\label{wingages}
\end{center}
\end{figure}

This is a significant problem because the wings in NGC~326 are at
least 100~kpc long; some XRGs have even longer wings and are in cooler
atmospheres.  Any expansion mechanism that depends on the sound speed
requires very old wings.  Figure~\ref{wingages} shows the ages implied
by (projected) wing lengths as a function of average expansion
velocity for several XRGs with detected IGM/ICM emission.  For
reference, it would take $\sim$200~Myr to inflate the wings of NGC~326
transonically and 250~Myr to do so buoyantly. 

Another possibility is that the radio galaxy is born in a dense
elliptical environment with the jet aligned with the major axis of the
atmosphere.  The cocoon formed by the spent jet material may then
become overpressured if the jet advance is stalled by the dense
atmosphere, and at some point it will rupture along the minor axis,
thereby forming wings \citep{capetti02}.  Before pressure equilibrium is established,
the outflow will be supersonic; afterwards, it will be akin to the
buoyant backflow scenario.  NGC~326 is not currently overpressured
relative to its surroundings \citep[M01,][]{worrall00}, so despite the
local ICM conforming roughly to the geometry required in the
\citet{capetti02} model, expansion would currently be driven by
buoyancy and bulk flows.  As described above, this expansion would be
subsonic.  

The circumstantial evidence therefore favors a reorientation scheme.
Detection of a bow shock ahead of the wings would be a strong
indictment of the backflow models, but there are insufficient counts
in the CXO image to perform an adequate search.  We note that the
temperature map in Figure~\ref{tempmap} reveals bins to the northeast
and northwest of the cavity that are $1.5-2$~times hotter than the
surrounding bins, and extraction of (low signal) spectra from bins
half the size ($\sim$700~counts) suggests the ICM is indeed hotter
closer to the cavity edges.  About four times as many counts are
required to determine if this is shock heating. 

On the other hand, the problem with reorientation models, as noted by
W95 and \citet{gopal03}, is that the wings have Z-shaped morphology
and the eastern wing has an off-axis extension to the northwest
(Figure~\ref{radio_maps}).  In response to the Z-shaped wings, \citet{gopal03} proposed a 
hybrid hydrodynamic and reorientation model in which a stream of ISM from a
merging galaxy deflects the jet, thereby forming a ZRG.  If a SMBH
binary is formed by this merger, upon coalescence the jet axis will
change and the ZRG becomes a fossil; the new lobes give the overall
radio emission X-shaped morphology.  This model was proposed for
NGC~326 by \citet{gopal03} and updated for the same by \citet{zier05}
with the suggestion that a stream of dense gas bent the jets at a
radius between 25--50~kpc from the core.  For reference, the projected
separation between the nuclei of the dumbbell components in NGC~326 is
$\sim$6.4~kpc.  

A strength of the \citet{gopal03} model is that it naturally 
explains the tendency for XRGs to fall near the FR~I/II break
\citep[with the most recent confirmation in][]{cheung09}.  This is because
deflecting a jet requires (from purely hydrodynamic considerations)
the pressure exerted by the jet on the ISM to be balanced by the
pressure of the gas.  For reasonable estimates of the density and bulk
motion of streams of merging gas, highly relativistic jets cannot be
stopped.  Indeed, the \textit{non-winged} FR~II sources
whose jets are deflected by interaction with dense gas also tend to
have radio luminosity close to the FR~I/II divide
\citep[e.g. PKS~2153$-$69;][]{young05} or hybrid lobe morphology
\citep[e.g. 3C~321;][]{evans08}.

However, a jet weak enough to be deflected by the ram pressure
of the ISM would likely inflate lobes at some fraction or low multiple of 
the sound speed (as is thought to be the case with FR~I sources). 
Tying expansion to the sound speed suggests very long
radio galaxy lifetimes (Figure~\ref{wingages}).  Apart from fueling
the AGN steadily, this timescale would require the bending agent to be
in contact with the jet at favorable orientations on both sides of the
galaxy for up to hundreds of Myr.  A gas stream 25--50~kpc from the
AGN with bulk motion of $\sim$200~km~s$^{-1}$ \citep[comparable to
  streams in similar systems;][]{zier05} would travel perhaps tens of
kpc in this time, meaning the stream must be very long and fortunately
oriented in order to keep the jet bent at roughly the same angle at
the same radius from the core.  Although conceivable, this is a very
special circumstance.  Another objection in the same vein is that the
time between the production of these streams of rotating gas and the
coalescence of a SMBH binary that engenders the present X-shaped
morphology must also be up to hundreds of Myr or the wings would not
have gotten so long, regardless of the stream geometry.  The
time required for a SMBH binary to coalesce is unknown, but one might
imagine that if the radio galaxy were driven by a component of the
SMBH binary over 100~Myr, accretion from the same disk of gas would
cause the spins to become aligned prior to the merger and forestall a
spin-flip upon coalescence \citep{bogdanovic07}.  It is possible that
the \citet{gopal03} model is essentially correct, but it faces two
hurdles: there is presently no evidence for the required streams in
NGC~326 and the streams, if found, must be able to bend a powerful
jet.

It is possible that the Z-shaped morphology formed after
reorientation.  The buoyant backflow and overpressured cocoon models
both take advantage of the anisotropy of the atmosphere to push the
radio lobes in the right direction.  This motion would still occur
with fossil relics.  Suppose that, while active, the lobes of the
wings formed an axis across the nucleus, but since the extinction of
the jet in that direction evolve buoyantly.  If the major axis of the
pocket of ICM they inhabit is somewhat off-axis from the major axis of
the radio galaxy \citep[but still close enough to satisfy
  the][correlation]{capetti02}, buoyancy will tend to lift the wings
away from the higher pressure core in opposite directions.  The
centers of the wings are about 25~kpc away from a line that would
connect them to the nucleus, implying a rise time of less than
50~Myr.  This is consistent with the age of the active lobes.  There
is evidence that the wings \textit{do} respond to the pressures in the
ICM: both appear to be following a ridge of brighter ICM emission
(Figure~\ref{hardness_compare}), forming a large but shallow arc.
Assuming the wings were bent into this shape after the jet ceased, the
radius of curvature compared to a straight line through the wings
suggests buoyancy would produce this morphology within 40--70~Myr
depending on the distance from the core (this estimate does not take
into account spatial variations in the buoyant terminal velocity due to an
anisotropic medium).  
However, the plume to the northwest of the eastern wing
(Figure~\ref{radio_maps}) would take longer to rise and is not well
explained by this scenario.

\subsection{Wings and Radio-Mode Feedback}

In any plausible wing formation scenario, the very long wings of
NGC~326 are presently evolving passively in the ICM, i.e., as buoyant
structures subject to hydrodynamic forces.  They are therefore akin to
``dead'' radio galaxies \citep{reynolds02} in which the jet has ceased \citep[in the
  model of][this is precisely what they are]{merritt02}.  Dead radio
galaxies are notoriously rare \citep[for one example,
  see][]{worrall07}, but crucial to the paradigm of ``radio-mode''
AGN feedback.  

In radio-mode feedback, radiative cooling in the ICM is regulated by
the condensation of gas onto the central galaxy, fueling an AGN that
drives energetic outflows into the ICM, heating it.  X-ray cavities
associated with radio lobes provide incontrovertible evidence that
these outflows exist, interact with the ICM by inflating bubbles of
relativistic plasma in it, and deposit sufficient energy to regulate
cooling \citep[for a recent review, see][]{gitti11}.  However, it is
not clear that these outflows \textit{do} regulate cooling, since we
do not know how this energy is thermalized and isotropized or on what
timescales this occurs. 

Several heating mechanisms have been identified.  These include 
\textit{active} heating through strong
or weak shocks; adiabatic expansion of a jet-blown cavity; and
damped sound waves associated with the cavity inflation, and
\textit{passive} heating such as entrainment of material in a
buoyant plume or infall of displaced gas behind a buoyant bubble
\citep[][and references therein]{mcnamara07}.  Although X-ray
observations demonstrate that all of these occur, their relative
importance (which may be a function of time, environment, and jet
power) is unknown.  A key problem is that the life cycle of AGNs (which
itself may be a function of redshift and whether the host galaxy is in
a group or a cluster) is not understood.  For example, an AGN with a
short duty cycle punctuated by brief, powerful outbursts may be less
effective at heating than an AGN that more or less continually blows
small bubbles \citep[``effervescent'' heating;][]{begelman01}.  Dead
radio galaxies are therefore of interest, since their frequency is a
measure of the AGN duty cycle and the rate at which they heat the ICM
is a measure of the importance of passive heating.  

Thus, wings may be of interest as probes of feedback.  If XRG wings
are dynamically similar to dead radio galaxies, they offer a few
additional advantages as laboratories of dead sources.  First, dead
radio galaxies are most frequently detected as ``ghost'' cavities
unassociated with GHz emission
\citep[e.g.][]{fabian00,rafferty06,dong10}, but ghost cavities are
typically detected near the group or cluster core where the surface
brightness is high.  It can therefore be difficult to distinguish
between features associated with old outbursts and the active ones.
Wings, on the other hand, can extend more than 100~kpc from the AGN
and in a very different direction.  Second, it is difficult to age
ghost cavities, whereas a better understanding of wing formation would
yield the ages of both the active lobes and wings in XRGs.  Third,
winged sources are typically located in group rather than cluster
environments.  Until recently, feedback in groups has been largely
neglected, but non-gravitational processes are even more important in
these systems than in clusters, and the total number of galaxies in
groups far outweighs the cluster population. 

Our examination of the formation history of NGC~326 suggests another
advantage to XRG wings: if the reason XRGs are typically found near
the FR~I/II break because they represent decaying outbursts, the total
active and passive energy injection rates over the history of the
source can be measured in the same system.  The deep X-ray
observations required for this kind of measurement only exist in a
handful of clusters with both active lobes and ghost cavities \citep[the
most prominent, of course, is Perseus;][]{fabian06}.  Since groups are
much fainter, it is essential to have good targets.  

The CXO observation of NGC~326 is not deep enough to carry out heating
measurements, but the surface brightness of features detected suggests
that an exposure with four times the counts (twice the signal) would
suffice.  For example, the high-temperature front has about
4,000~counts in its vicinity.  To determine if it is a shock and
measure its strength to 10\% accuracy requires surface brightness
values and deprojected temperatures to at least this accuracy in
semi-annular bins about 15~arcsec wide at a radius of 50~kpc from the
nucleus.  This would require about 15,000~counts per bin.  Since
strong shocks are much more efficient at irreversibly heating the ICM
than weak ones, the shock strength is of interest.  

Likewise, the cavity in the wing would be detected at the $7\sigma$
level in an exposure of this duration, and cavities associated with
the other lobes (which we assume to be there) would be easily
detected, barring pathological projection effects.  In the wing, this
quality of detection would place tight constraints on the cavity size,
important for determining the cavity stability and placing boundaries
from which to measure the amount of entrained material, uplifted
material, and temperature or metallicity gradients.  An image with
twice the signal would allow for adapative binning at scales
significantly smaller than the lobe size with temperature measurements
accurate to 20\%.  Specific entropy (a surer measure of irreversible
heating than temperature) could also be measured on these scales.
Metallicity could be determined in slightly larger
bins that would still be smaller than the lobe size, enabling a
measurement of metallicity gradients around and behind the cavity.
Given an accurate measurement of the terminal buoyant velocity of the
bubble, the displacement heating rate \citep{birzan04} could then be
measured.  

Thus, while the extant CXO image lacks the depth to take advantage of
the measurements sketched here, it demonstrates that such measurements
are well within reach.  The utility of XRG wings depends on
high-quality X-ray data and better understanding wing origin, so flux
estimates in other systems are important.  Only 13~XRGs have CXO or
XMM-\textit{Newton} data, and only $\sim$8 of these have significant
IGM/ICM emission \citep[model flux estimates provided
  in][]{hodges-kluck10a}.  Only two of these have moderately deep
exposures (NGC~326 and 4C~+00.58).  The CXO image presented here
suggests that for larger XRGs, the higher effective area of
\textit{XMM-Newton} makes it a better instrument for these types of
measurements.  

\section{Summary and Conclusions}

We report on a 100~ks CXO image of NGC~326, an XRG with wings 100~kpc
long that resides in a poor cluster.  Owing to its dramatic wings,
NGC~326 has frequently been the subject of studies attempting to
explain wing origin
\citep[W95;][]{wirth82,merritt02,gopal03,dennett02,zier05,capetti02,lal05}.
The CXO image provides an important step forward in evaluating these
hypotheses due to the identification of several X-ray features
associated with the radio galaxy.  These include a high-temperature
front to the southeast of the radio galaxy (Figure~\ref{tempmap}), which we assume to be a
shock front, high-temperature knots of emission around the radio
galaxy akin to cavity rims (Figure~\ref{hardness_compare}), and a cavity associated with the eastern
wing (Figure~\ref{cavity}).  The high angular resolution of CXO compared
to ROSAT also allows us to remove point sources and characterize the
broader ICM, revealing an X-ray peak associated with the northern lobe
of the radio galaxy rather than the ISM and an elliptical enhancement
around the radio galaxy \citep[first identified
  in][]{hodges-kluck10a}.  

By characterizing the immediate environment of the radio galaxy and
taking into account spectral aging of the radio emission in the
wings, we can construct a reasonable history for the AGN outbursts.
The jets, lack of hot spots, and low radio power
($10^{41}$~erg~s$^{-1}$) indicate a weak FR~I-like source today, but 
the putative shock front, lobe size and morphology, and GHz emission at the wing tips
indicate a powerful FR~II source no more than $\sim$30~Myr ago.  We
therefore suggest the kinetic jet power could be described as 
$P_{\text{jet}} = 10^{45} e^{-t/t_0}$~erg~s$^{-1}$, where $t_0 =
10$~Myr for the current outburst.  This function would account for the
observed features as well as produce about the right amount of cavity
enthalpy.  The energy required to
inflate cavities the size of the lobes places the
current kinetic jet power between $5\times 10^{42} - 5\times
10^{43}$~erg~s$^{-1}$.  The radio power suggests that the flow is
therefore still radiatively inefficient.  

NGC~326 is the best studied XRG, and the variety of wing formation
models associated with it is a testament to the difficulty of proving
any one scenario. 
The sound speed in the ICM is only 500~km~s$^{-1}$
($0.5$~kpc~Myr$^{-1}$), implying that the long (100~kpc) wings must have
expanded supersonically.  This disfavors the backflow models, which
have yet to convincingly demonstrate that internal lobe flows can
generate sustained supersonic expansion; simulations indicate that
they become turbulent and dissipate their thrust before reaching the
wing edges.  In fact, any model in which the wing growth is related to 
the sound speed is disfavored in systems with long wings
(Figure~\ref{wingages}).  The most likely alternative is that the
wings are fossil relics illuminated by the present outburst, but
\citet{gopal03} note that a simple spin-flip would not explain the
Z-shaped morphology of the wings.  However, their model relies on
bending a moderately weak jet, which would not be expected to propel
hypersonic lobes.  As fossil radio lobes would be subject to
hydrodynamic forces, it is possible that the Z-shaped morphology of
the wings is produced by buoyant evolution in an anisotropic
atmosphere.  It would take less than 50~Myr for the wings to rise
buoyantly away from an axis connecting them through the nucleus to
their present-day location, which is consistent with the age of the
active lobes and the age required for the wings to respond to 
the anisotropy of the large-scale cluster emission to form the
understated C-shaped arc (Figure~\ref{hardness_compare}).  However,
this scenario fails to explain the plume on the eastern wing.  It is
also notable that the tendency for XRGs to have jets co-aligned with
the major axes of their host galaxies \citep{capetti02,saripalli09}
and hot atmospheres \citep{hodges-kluck10a} extends to XRGs with very
long wings.  This is unexplained by any reorientation model, and
remains compelling evidence that the environments of most XRGs help
determine their morphology.  

XRGs provide an important target for characterizing AGN feedback
through the study of ``ghost'' cavities, both as pathfinders (they
illuminate the cavities in the GHz bands) as well as targets which are
readily distinguished from the active outburst.  The heating rate
associated with dead outbursts can then be measured in isolation, and
compared to the timescales required to thermalize and isotropize the
energy in the ICM.  In cases such as NGC~326, where we believe most of
the energy associated with the active outburst has already been
deposited, we can use the AGN lifetime to estimate an average heating
rate, assuming some duty cycle.  XRGs are therefore interesting beyond
the origin of their bizarre morphology.  

\acknowledgments

The authors thank the anonymous referee for helpful comments.
CSR thanks support from the National Science Foundation under grant AST0908212.

{\it Facilities:} \facility{CXO}, \facility{VLA} 


\end{document}